**Footprints of ancient balanced polymorphisms in genetic variation data**

Ziyue Gao[*,1], Molly Przeworski[§, †, ‡] and Guy Sella[‡,2]

[*] Committee on Genetics, Genomics and Systems Biology, University of Chicago, Chicago, IL 60637

[§] Dept. of Human Genetics, University of Chicago

[†] Howard Hughes Medical Institute, University of Chicago

[‡] Dept. of Ecology and Evolution, University of Chicago

---

[1] *Corresponding author*
[2] *Present address*: Dept. of Biological Sciences, Columbia University, 740A Mudd, M.C. 2435, New York, NY 10027

ABSTRACT

When long-lived, balancing selection can lead to "trans-species" polymorphisms that are shared by two or more species identical by descent. In this case, the gene genealogies at the selected sites cluster by allele instead of by species and, because of linkage, nearby neutral sites also have unusual genealogies. Although it is clear that this scenario should lead to discernible footprints in genetic variation data, notably the presence of additional neutral polymorphisms shared between species and the absence of fixed differences, the effects remain poorly characterized. We focus on the case of a single site under long-lived balancing selection and derive approximations for summaries of the data that are sensitive to a trans-species polymorphism: the length of the segment that carries most of the signals, the expected number of shared neutral SNPs within the segment and the patterns of allelic associations among them. Coalescent simulations of ancient balancing selection confirm the accuracy of our approximations. We further show that for humans and chimpanzees, and more generally for pairs of species with low genetic diversity levels, the patterns of genetic variation on which we focus are highly unlikely to be generated by neutral recurrent mutations, so these statistics are specific as well as sensitive. We discuss the implications of our results for the design and interpretation of genome scans for ancient balancing selection in apes and other taxa.

INTRODUCTION

Balancing selection is a mode of adaptation that leads to the presence of more than one allele in the population at a given time. Although balancing selection is commonly assumed to be the result of heterozygote advantage (also known as over-dominance), the possible sources of balancing selection are more diverse and include negative frequency-dependent selection and temporally or spatially heterogeneous selection (LEVENE 1953; NAGYLAKI 1975; WILSON and TURELLI 1986; FISHER 1990). The common outcome of these diverse modes of balancing selection is the persistence of genetic variation beyond what is expected from genetic drift (or directional selection) alone (DOBZHANSKY 1970; KIMURA 1983).

Balancing selection leaves discernible footprints in genetic variation data (CHARLESWORTH 2006). In particular, the effects of old balancing selection on variation data within a species are well understood (*e.g.*, HUDSON and KAPLAN 1988; KAPLAN *et al.* 1988). A site under balancing selection has a deeper genealogy than expected under neutrality, with long internal branches. Because of linkage, the genealogies at nearby sites will have similar properties. As a result, balancing selection leads to higher diversity and more intermediate frequency alleles at linked neutral sites. These considerations suggest that targets of old balancing selection could be identified from their footprints in genetic variation data (*e.g.*, BUBB *et al.* 2006; ANDRÉS *et al.* 2009). A challenge, however, is that these footprints can also be produced by neutral processes alone and are thus not specific, raising the concern of a high false discovery rate (KREITMAN and DI RIENZO 2004).

When balancing selection is sufficiently long-lived that it predates the split of two or more species, it may lead to a "trans-species polymorphism" that is shared by two or more species identical by descent (FIGUEROA *et al.* 1988). If the species diverged long enough ago that no shared variation is expected by chance, then the persistence of ancestral variation to the present in both species is a distinctive signature of balancing selection, which is unlikely to be generated by neutral processes (WIUF *et al.* 2004; CHARLESWORTH 2006). For that reason, a trans-species polymorphism is considered the least equivocal evidence for ancient balancing selection (CHARLESWORTH 2006).

Only a handful of convincing examples of trans-species polymorphisms have been reported so far: the major histocompatibility complex in vertebrates, the opsins in New World monkeys, self-incompatibility genes in plants, the ABO blood groups in primates, and more recently, several non-coding regions in humans and chimpanzees (FIGUEROA *et al.* 1988; IOERGER *et al.* 1990; SURRIDGE and MUNDY 2002; SEGUREL *et al.* 2012; LEFFLER *et al.* 2013). On this basis, ancient balancing selection is thought to be rare. Moreover, most trans-species polymorphisms found to date are involved in self/non-self recognition, potentially suggesting that such long-lived balancing selection only occurs in highly specific circumstances. Alternatively, it is possible that ancient balancing selection is more widespread than appreciated but remains largely unrecognized. Indeed, trans-species polymorphisms are difficult to detect without the dense genome-wide variation data that have only recently become available (BUBB *et al.*). Moreover, the effects of trans-species polymorphism remain poorly characterized, so its footprints may have gone undetected in some instances (SEGUREL *et al.* 2012).

Since for species that are not too closely related, a trans-species polymorphism is unlikely to be due to chance, targets of ancient balancing selection can be identified by scanning genetic variation data from two or more species for shared polymorphisms (polymorphisms at homologous positions with the exactly same segregating alleles) (*e.g.*, ASTHANA *et al.* 2005). A major challenge, however, is that most shared single nucleotide polymorphisms (SNPs) between species are expected to be due to recurrent mutations, *i.e.*, the independent occurrences of the same mutation in both species. In such cases, although the resulting shared polymorphism mimics a trans-species polymorphism, the underlying gene genealogy is markedly different from that of a trans-species polymorphism. The unusual depth and distinctive topology of the genealogies around a trans-species polymorphism and the patterns of genetic variation that they generate can therefore help to distinguish these two cases, *i.e.*, to identify patterns that are both sensitive and specific to trans-species polymorphisms.

Motivated by these considerations, we focus on summaries of genetic variation data that capture key features of the underlying genealogical structure around trans-species polymorphism. We derive their distribution in order to help guide scans for trans-species polymorphisms and assess how helpful they are expected to be in identifying targets. Specifically, we consider a trans-species polymorphism that arises from balancing selection at a single site and describe the genealogical structures expected at linked sites. We then derive approximations for three summary statistics that reflect distinct aspects of this genealogical structure and together are sensitive and specific to the presence of a trans-species polymorphism: the distribution of the length of the segment that carries the signals, the expected number of shared SNPs and the expected linkage disequilibrium (LD) pattern among them. The accuracy of the approximations is confirmed by coalescent simulations. While in principle our derivations can be applied to any pair of related species, we focus on humans and chimpanzees, because this pair of species is well suited for scans for ancient balancing selection (see below); moreover, multiple examples of trans-species polymorphisms have already been found in this species pair (LEFFLER *et al.* 2013), and we can interpret their evolutionary history in light of our results.

MODEL AND RESULTS

**The model**

We consider a simple demographic scenario in which an ancestral species splits into two species $T$ generations ago and there is no subsequent gene flow between them. For simplicity, we assume panmixia and constant population sizes for the ancestral ($N_a$) and descendant species ($N_e$). We consider a bi-allelic ancestral polymorphism under strong balancing selection, such that allele $A1$ is maintained at constant equilibrium frequency $p$ and allele $A2$ at $q=1$-$p$, and assume the two alleles reach their equilibrium frequencies immediately after the balanced polymorphism arose. In our model, there are no subsequent mutations at the selected site, so there is no allele turnover and, consequently, all chromosomes carrying the same allele (even from two different species) are identical by descent. When the ancestral species splits into two, this selected polymorphism passes into each of the descendant species and is maintained by balancing selection at constant equilibrium frequency until the present, resulting in a polymorphism shared between species.

Throughout, we consider $T$ much larger than $N_e$. This condition is needed for trans-species polymorphisms to be a clear-cut manifestation of balancing selection, as otherwise a substantial proportion of the trans-species polymorphisms in the genome will be due to neutral incomplete lineage sorting (WIUF *et al.* 2004). Indeed, a neutral trans-species polymorphism will occur if three conditions are met: 1) at least two lineages from each species do not coalesce by the time of the split (throughout the paper, we measure time backwards, unless otherwise specified.) 2) the order of coalescent events in the ancestral species is the right one (*i.e.*, the first coalescent event regarding the four lineages is between lineages from different species); 3) there is a mutation on the appropriate lineage(s) on the genealogy. Under our demographic assumptions, the probability of a neutral trans-species polymorphism is mainly determined by the probability of the first condition, which is $(e^{-T/2Ne}) \times (e^{-T/2Ne})$. This simple derivation indicates trans-species polymorphism is highly unlikely if the species split a sufficiently long time ago (measured in units of $2N_e$ generations). An implication is that at a neutral site unlinked to a site under balancing selection, the local gene tree should cluster by species (Figure 1A).

**The genealogies of sites around a balanced trans-species polymorphism**

We model the coalescent process at sites around a balanced trans-species polymorphism by using the framework of Hudson and Kaplan (HUDSON and KAPLAN 1988). The Hudson-Kaplan model is a form of structured coalescent (NORDBORG 1997), in which allelic classes at the selected site are analogous to subpopulations: sequences carrying the same allele are exchangeable, whereas two sequences carrying different alleles cannot coalesce unless one of the sequences migrates into the allelic group of the other by undergoing between-class recombination; henceforth, we assume all recombination to be crossing-over (ignoring gene conversion without exchange of flanking markers) and use "recombination" to denote "between-class recombination" unless otherwise specified.

The gene trees at neutral sites change with the genetic distance from the selected site (Figure 1A). At a tightly linked site, the tree has the same topology as the selected site, notably lineages carrying the same selected allele coalesce before lineages carrying different selected alleles and the tree clusters by allele instead of by species (the orange topology in Figure 1A). We term the segment with this topology the *ancestral segment*. At a site a little farther from the selected site, the probability of recombination between this neutral site and the selected one is larger, so a recombination event may occur in one of the species before the split time. Assuming, without loss of generality, that the recombination event takes place in species 1, all lineages in species 1 would carry the same allele after the recombination and thus are closely related to each other, but this is not true for species 2: some lineages from species 2 are more closely related to lineages from species 1 than to the other lineages from species 2 (the yellow topology in Figure 1A). At a site farther from the selected site, recombination events may occur in both species before the split, so the tree reflects the species relationship; however, the coalescent time in the ancestral population varies dramatically, depending on the allelic identities of the two ancestral lineages (the green and blue topologies in Figure 1A).

A natural way of examining whether a shared polymorphism is identical by descent is therefore to build a phylogenetic tree of haplotypes from both species around the selected site and test if there is strong support for a tree that clusters by allele. But over what window size? The window must include the information in the ancestral segment; yet, importantly, it cannot be too large or it will also incorporate segments with other topologies that will dilute the signal (Nordborg and Innan 2003). It is also important to understand which features of the data are indicative of a topology that clusters by allele. Among these are an absence of fixed difference between species, the presence of neutral shared SNPs in addition to the selected one and the allelic associations among shared SNPs (with the same alleles coupled in the two species).

Motivated by these considerations, we ask: (i) What is the *distribution of the length* of the ancestral segment for a sample of four chromosomes, one from each allelic class in each species? (ii) What is the *expected number of neutral shared polymorphisms* in the ancestral segment for that sample? (iii) What are the *expected number of neutral shared polymorphisms* and *the linkage disequilibrium (LD) patterns* among them, for a sample of more than four chromosomes? Together, answers to these questions inform the scale over which to test for the presence of a trans-species polymorphism and the strength of the expected signals.

## Coalescent simulations of a balanced polymorphism

Throughout this article, we use coalescent simulations to evaluate the accuracy of the analytic approximations. The simulations are done using the coalescent program *ms_sel*, an extension of the program *ms* (HUDSON 2002), kindly provided to us by R. Hudson. The program generates samples of chromosomes carrying a selected polymorphism, allowing for an arbitrary trajectory of the selected alleles; it can output the gene tree of each non-recombining segment or the simulated sequences under an infinite sites mutation model. We consider the same model as in our analytic derivation: two species have the same constant effective population size ($N_e$) until they merged $T$ generations ago to form an ancestral species with a constant effective population size ($N_a$), and a bi-allelic polymorphism is maintained at a constant equilibrium frequency $p$ immediately after the derived allele arises $T_S$ generations ago. We note that since the program outputs haplotypes rather than genotypes, we are implicitly assuming that there is no phasing error.

**The length of the ancestral segment:** For each combination of parameters ($N_e$, $N_a$, $T$ and $p$), we run 1,000 replicates, each of which contained four (or three) chromosomes of length 0.012 cM (or on average 10 kb in humans) centered on the balanced polymorphism. (The length of the simulated segment and the position of the balanced polymorphism are the same in all simulations, unless otherwise specified.) For the case with four chromosomes, the sample consists of one chromosome from each allelic class in each species; for the case with three chromosomes, (without loss of generality) two are from the two allelic classes in species 1, and the other one is from *A1* class in species 2. The gene trees are output with the option –T in *ms_sel* and the topologies and coalescent times were analyzed with the R package "ape" (Paradis *et al.* 2004). To obtain the contiguous segment length in each replicate, we sum up the lengths (in genetic distance) of all consecutive segments with a topology that clusters by allele on each side of the selected locus.

**The expected number of shared SNPs:** Our approximation of the expected number of neutral SNPs shared between species relies on estimation of the expected coalescent time between two lineages carrying different selected alleles in the ancestral species, so we just simulate the coalescent process in the ancestral species. We run 10,000 replicates for each combination of parameters ($N_a$, $T_S$ and $p$). Each replicate simulates two chromosomes in the ancestral species, one from each allelic class. We proceed as above to record the coalescent time for each non-recombining segment.

**The influence of sample size on the number of shared SNPs and the linkage disequilibrium (LD) patterns among them:** We consider the simple demographic model with a split time of $T=20N_e$ generations before present. Given that the age of a trans-species polymorphism does not affect the LD between neutral and selected shared SNPs in extant samples, we assume an extremely old age of the balanced polymorphism ($T_S=400N$) in order to increase the probability of observing at least one neutral shared SNP in a replicate. We run 100,000 replicates, each of which contains 50 chromosomes from each allelic class in each species. We then randomly sample the same number of chromosomes (n=2, 4,10, 20, 50) from the two species and use an R script to identify SNPs shared between species from the sampled data. Since the *ms_sel* program assumes an infinite sites mutation model, every shared SNP in the sample is identical by descent. To ensure independence of our observations, we characterize the LD between a neutral SNP and the selected one by picking one neutral shared SNP at random from each replicate.

**The length of the ancestral segment**

We first derive the length of the ancestral segment, corresponding to the scale over which there are no fixed differences between species and there may be shared neutral SNPs, *i.e.*, the scale over which we expect the clearest evidence for a trans-species polymorphism. Its length is determined by the recombination events on the genealogy in stages I and II defined in Figure 2. We begin by considering the length of the ancestral segment for a sample of four chromosomes, one from each allelic class in each species (stage I does not exist in this case). In the Supporting Information, we show that so long as $T \gg N_e$, the effects of complex recombination events in the history of the sample can be neglected. In other words, the length of this segment is well approximated by the length of the segment in which no recombination event occurs until all the lineages from each allelic class coalesce into their common ancestor.

The duration of stage II includes the split time and the time ($t$) required for lineages carrying the same allele to coalesce in the ancestral species (Figure 2). Denoting the coalescent times for the two *A1* lineages and the two *A2*s by $t_1$ and $t_2$ respectively:

$$t = max\{t_1, t_2\},$$

$$t_1 \sim Exp(\frac{1}{2N_a p}),$$

$$t_2 \sim Exp(\frac{1}{2N_a q}).$$

Therefore, the probability density of $t$ is:

$$f_t(u) = \frac{1}{2N_a pq}(pe^{-\frac{u}{2N_a q}} + qe^{-\frac{u}{2N_a p}} - e^{-\frac{u}{2N_a pq}}). \qquad (1)$$

On each side of the selected site, the length of the ancestral segment is well approximated by the distance to the nearest recombination event in stage II (Figure 2). We denote this (genetic) distance by $X$ and the genetic distance to the nearest recombination on each lineage by $x_1^{A1}$, $x_1^{A2}$, $x_2^{A1}$ and $x_2^{A2}$ respectively, so that $X = \min\{x_1^{A1}, x_1^{A2}, x_2^{A1}, x_2^{A2}\}$. To calculate the distribution of $X$, we rely on the fact that the distance to the nearest recombination on each lineage is exponentially distributed:

$$x_1^{A1}, x_2^{A1} \sim Exp(q(T+t)),$$

$$x_1^{A2}, x_2^{A2} \sim Exp(p(T+t)).$$

If we assume that $x_1^{A1}$, $x_1^{A2}$, $x_2^{A1}$ and $x_2^{A2}$ are independent of each other, the conditional distribution of $X$ given $t$ can therefore be approximated by:

$$X \,/\, t \sim Exp(2(T+t)). \qquad (2)$$

In reality, the two lineages that coalesce first share an ancestor in the time period between $t_1$ and $t_2$ and thus are not independent. However, Equation (2) only slightly under-estimates $X$ (confirmed by simulations), because we over-estimate the reduction in the ancestral segment in the time period between $t_1$ and $t_2$ by about a third, and when $N_a \ll T$, that time period is negligibly short compared to the total length of stage II. The assumption that $N_a \ll T$ will often be satisfied, because for most species $N_a$ will be on the same order of magnitude as $N_e$, which is assumed to be much smaller than $T$. For simplicity, in what follows we rely on the approximation in Equation (2).

Combining Equations (1) and (2), we obtain the probability density function of $X$ as:

$$f_X(x) = \int_0^\infty f_{X|t}(x \mid u) f_t(u) du = \int_0^\infty 2(T+u) e^{-2(T+u)x} \frac{1}{2N_a pq} (pe^{-\frac{u}{2N_a q}} + qe^{-\frac{u}{2N_a p}} - e^{-\frac{u}{2N_a pq}}) du$$

$$= \frac{1}{N_a pq} e^{-2Tx} [p(\frac{T}{B} + \frac{1}{B^2}) + q(\frac{T}{C} + \frac{1}{C^2}) - (\frac{T}{D} + \frac{1}{D^2})], \qquad (3)$$

where $B = 2x + \dfrac{1}{2N_a q}$, $C = 2x + \dfrac{1}{2N_a p}$ and $D = 2x + \dfrac{1}{2N_a pq}$. When $p=q=0.5$, equation

(3) can be simplified to:

$$f_X(x) = \frac{4e^{-2Tx}}{(2N_a x + 1)(2N_a x + 2)}(T + \frac{N_a}{2N_a x + 1} + \frac{N_a}{2N_a x + 2}).$$

The distribution is insensitive to changes in the frequencies of the selected alleles (Table S1), because the shape of the distribution is primarily determined by recombination events that occur before the species split, the rates of which do not depend on $p$.

Since recombination events on the two sides of the chromosome are independent, the length of two-sided ancestral segment is the sum of the one-sided lengths on both sides, and the distribution of the two-sided length is given by the convolution of the one-sided distribution with itself. Considering parameters appropriate for humans and chimpanzees, we find that the predicted distribution of the length agrees extremely well with the simulation results (Figure S1, Table S1).

Our result for the length of the ancestral segment has important implications for genome scans of ancient balancing selection. Specifically, Equation (3) shows that the length of the ancestral segment shrinks exponentially with $T$. For instance, assuming plausible parameters for humans and chimpanzees ($N_e$=10,000, $N_a$=50,000, $T$=250,000 generations, $p$=0.5 and an average recombination rate of $r$=1.2×10$^{-8}$ per bp per generation), the expected length of the ancestral segment on one side of the selected site is 131 base pairs (bp) and the 95% quantile is about 400 bps. In contrast, for *Drosophila melanogaster* and *Drosophila simulans* (analyzed *e.g.* in Langley *et al.* 2012; Bergland 2013), the ancestral segment to one side of a single ancient balanced polymorphism will only be two base pairs on average and the upper 95-tile six base pairs (assuming $T=2\times10^7$, $N_a=10^6$, $p=0.5$ and $r=1.2\times10^{-8}$). This indicates that scans for ancient balancing selection would have little power for species with too old a split time. However, as we have seen, the split time should be sufficiently old for there to be little or no incomplete lineage sorting by chance. Thus, $T$ (in generations) must be much greater than $N_e$ but much smaller than $1/r$. In this regard, human and chimpanzee are a particularly well-suited pair, with a split time of approximately $18N_e \ll 1/r \approx 8\times10^7$ (WIUF *et al.* 2004).

By a similar approach, we also derive the length distribution of segments containing the orange or yellow topologies (see Supporting Information). In the yellow segments, there is recombination before the split time in one of the species, so the two lineages in that species are most closely related to each other. However, in the other species, two sequences carrying different selected alleles do not experience recombination before the split time, so they are on average more divergent than a pair carrying the same selected allele from different species. This unusual divergence pattern is also a signature of trans-species polymorphism but is less specific (because it is somewhat more likely to occur by chance; Figure 1C, also see SEGUREL *et al.* 2012).

**The number of shared SNPs in the ancestral segment**

SNPs shared between species can be identical by descent or be generated by recurrent mutations. Therefore, observing a single shared SNP does not in itself provide compelling evidence for ancient balancing selection. Patterns of genetic variation in the ancestral segment, however, can provide more specific evidence for trans-species balanced polymorphism (LEFFLER *et al.* 2013). Notably, this segment may carry neutral polymorphisms shared identical by descent in addition to the selected one. The expected number of such neutral shared SNPs at a given genetic distance from the selected site is proportional to the coalescent time between the two lineages remaining in stage III (Figure 2). In what follows, we therefore derive expressions for this coalescent time.

We begin by considering the limit of an infinitely old balanced polymorphism. After the samples from each allelic class have coalesced to a common ancestor in the ancestral species, the coalescent process of the remaining two lineages can be described following Hudson and Kaplan (HUDSON and KAPLAN 1988). The coalescent process at genetic distance *d* from the selected site can be described as an island model with two subpopulations, corresponding to the two allelic backgrounds. Let $T_1$ be the coalescent time for two lineages carrying the *A1* allele, $T_2$ for two lineages carrying the *A2* allele and $T_B$ for two lineages with different selected alleles. Conditioning on the outcome of the first step of the Markov process, we can write down the expected coalescent times in recursive equations:

$$E(T_1) = \frac{1}{\frac{1}{2qN_a} + 2pd} + \frac{2pd}{\frac{1}{2qN_a} + 2pd} E(T_B),$$

$$E(T_2) = \frac{1}{\frac{1}{2pN_a} + 2qd} + \frac{2qd}{\frac{1}{2pN_a} + 2qd} E(T_B),$$

$$E(T_B) = \frac{1}{pd + qd} + \frac{qd}{pd + qd} E(T_1) + \frac{pd}{pd + qd} E(T_2), \tag{4}$$

from which we obtain:

$$E(T_1) = 2N_a + \frac{2pN_a(2q-1)}{4pqN_ad+1},$$

$$E(T_2) = 2N_a + \frac{2qN_a(2p-1)}{4pqN_ad+1},$$

$$E(T_B) = 2N_a + \frac{1}{d}.$$

The expected number of shared neutral SNPs in the ancestral segment follows from the coalescent times. If we assume that the shared segment is $L$ base pairs long (not including the selected site), with a uniform recombination rate $r$ and a uniform mutation rate $\mu$, the expected number of shared neutral SNPs is obtained by summing over distances:

$$\mu \cdot 2\sum_{i=1}^{L} E(T_{B,ir}) = 2\mu \sum_{i=1}^{L} (2N_a + \frac{1}{ir}) = 4N_a\mu L + 2\frac{\mu}{r}\sum_{i=1}^{L}\frac{1}{i}.$$

When the balanced polymorphism is not infinitely old, the coalescent process differs before and after the balanced polymorphism arose. We assume the balanced polymorphism arose $T_S$ generation after all lineages in the same allelic group coalesced (*i.e.*, the end of stage II in Figure 2) and therefore divide the genealogical process into the selection phase and the neutral phase.

Because the coalescent process is not homogeneous in time, the expected coalescent times considered in Equation (4) change with time and thus the first-step analysis does not provide a simple solution. We therefore apply a different approach. The coalescent process at a neutral site $d$ Morgans away from the selected site can be described as follows. In the first $T_S$ generations, when selection is acting, the state of the sample of two lineages in generation $t$ is described by $Q(t)=(i, j)$, where $i$ is the number of lineages carrying the derived allele *A1* and $j$ is the number of lineages carrying the ancestral allele *A2*. Once the two lineages coalesce, we no longer care about the allelic identity, so we merge the two states (1,0) and (0,1) into one state, denoted by (*). The transition matrix ($P_S$) between the four states (1,1) (0,2), (2,0), and (*) is:

$$P_S = \begin{bmatrix} 1-d & qd & pd & 0 \\ 2pd & 1-2pd-\dfrac{1}{2N_a q} & 0 & \dfrac{1}{2N_a q} \\ 2qd & 0 & 1-2qd-\dfrac{1}{2N_a p} & \dfrac{1}{2N_a p} \\ 0 & 0 & 0 & 1 \end{bmatrix}.$$

After the first $T_S$ generations, when balancing selection no longer plays a role, there are two possible states, corresponding to two (**) and one remaining lineages (*) and the transition matrix is:

$$P_N = \begin{bmatrix} 1-\dfrac{1}{2N_a} & \dfrac{1}{2N_a} \\ 0 & 1 \end{bmatrix}.$$

The transition matrix between the selection and neutral phases is:

$$Q = \begin{bmatrix} 1 & 0 \\ 1 & 0 \\ 0 & 1 \\ 0 & 1 \end{bmatrix}.$$

The transition from (2,0) at the end of the selection phase to (*) at the beginning of the neutral phase comes from our assumption (forward in time) that the derived selected allele goes to frequency $p$ immediately after it arises, so backwards in time any two lineages carrying the derived allele have to coalesce at generation $T_S$.

The average coalescent time of interest is equal to the expected number of steps needed to enter state (*) starting from state (1,1). Note that the entries in the first row of the $k^{th}$ power of $P_S$ are the probabilities of being in the four states after $k$ steps, starting from (1,1). Denoting the four entries in the first row by $\Phi^{(k)} = (\Phi_{(1,1)}^{(k)}, \Phi_{(0,2)}^{(k)}, \Phi_{(2,0)}^{(k)}, \Phi_{(*)}^{(k)})$, the average coalescent time of interest, $E(T_B)$, is:

$$E(T_B) = (\Phi_{(1,1)}^{(T_S)} + \Phi_{(0,2)}^{(T_S)})(T_S + 2N_a) + \Phi_{(2,0)}^{(T_S)}T_S + \Phi_{(*)}^{(T_S)}E(T_B \mid T_B \leq T_S). \qquad (5)$$

In turn, $E(T_B/T_B{\leq}m)$ satisfies the recursion:

$$E(T_B \mid T_B \leq m+1) = \frac{\Phi_{(*)}^{(m)}}{\Phi_{(*)}^{(m+1)}} E(T_B \mid T_B \leq m) + \frac{\Phi_{(*)}^{(m+1)} - \Phi_{(*)}^{(m)}}{\Phi_{(*)}^{(m+1)}}(m+1),$$

from which we can re-write $E(T_B/T_B{\leq}m)$ as:

$$E(T_B \mid T_B \leq m) = \frac{\sum_{i=2}^{m}(\Phi_{(*)}^{(i)} - \Phi_{(*)}^{(i-1)})\cdot i}{\Phi_{(*)}^{(m)}}. \qquad (6)$$

Thus, for given parameters, we can calculate $\Phi^{(k)}$ numerically and then use Equations (5) and (6) to derive the expected coalescent time $E(T_B)$ (a *Mathematica* notebook that implements these calculations is provided in the Supporting Information).

Alternatively, a closed form approximation of $\Phi^{(k)}$ can be obtained by ignoring recombination once the process leaves state (1,1). This approximation will lead to an under-estimate of the coalescent time; however, it is expected to perform well on the length scale of interest, because the ancestral segment is so short that the recombination rate within it is on the order of $1/(2T)$, which is much smaller than the coalescence rate of $\sim 1/(2N_a)$ when $N_a << T$. In this approximation, the transition matrix during the selection phase $P_S$ simplifies to an upper triangular matrix:

$$P_S \approx \begin{bmatrix} 1-d & qd & pd & 0 \\ 0 & 1-\dfrac{1}{2N_a q} & 0 & \dfrac{1}{2N_a q} \\ 0 & 0 & 1-\dfrac{1}{2N_a p} & \dfrac{1}{2N_a p} \\ 0 & 0 & 0 & 1 \end{bmatrix}.$$

And with this matrix we can solve for $\Phi^{(k)}$ to obtain:

$$\Phi_{(1,1)}^{(k)} \approx (1-d)^k,$$

$$\Phi_{(0,2)}^{(k)} \approx \frac{qd[(1-d)^k - (1-\dfrac{1}{2N_a q})^k]}{\dfrac{1}{2N_a q} - d},$$

$$\Phi_{(2,0)}^{(k)} \approx \frac{pd[(1-d)^k - (1-\dfrac{1}{2N_a p})^k]}{\dfrac{1}{2N_a p} - d},$$

$$\Phi_{(*)}^{(k)} \approx 1 - (1-d)^k - \frac{qd[(1-d)^k - (1-\dfrac{1}{2N_a q})^k]}{\dfrac{1}{2N_a q} - d} - \frac{pd[(1-d)^k - (1-\dfrac{1}{2N_a p})^k]}{\dfrac{1}{2N_a p} - d}.$$

By substituting these expressions into Equation (5) and (6), we obtain an analytic approximation of the expected coalescent time $E(T_B)$.

In Figure 3, we show a comparison of the expected $T_B$ obtained from the numerical calculation, analytic approximation and coalescent simulations. We consider combinations of the parameters $N_a$, $T_S$, and $p$ that are plausible for the ancestral species of humans and chimpanzees. As expected, the numerical calculation predicts the mean coalescent time from the simulations very well, whereas the analytic approximation tends to slightly under-estimate the mean.

However, the approximation is much faster and easier to obtain than the numerical calculation and, on the scale of the ancestral segment, it gives similar results.

Based on these results, we can predict the expected number of neutral shared SNPs around a trans-species balanced polymorphism in humans and chimpanzees (Figure 4). As an illustration, assuming uniform mutation and recombination rates and the average recombination rate for humans, when the balanced polymorphism is 20 million years old from present, we expect on average about two additional neutral shared SNPs - more if the polymorphism is older. For a region where recombination is lower than the genome average, the ancestral segment is longer and could plausibly contain as many as a dozen neutral shared SNPs. In turn, for *D. melanogaster* and *D. simulans*, even though the contiguous ancestral segment is expected to be two base pairs in length only (assuming $T=2\times10^7$ generations, $N_a=10^6$, $p=0.5$ and $r=1.2\times10^{-8}$ per bp per generation), both sites can harbor shared polymorphisms if the balanced polymorphism is sufficiently old (assuming that the balanced polymorphism is infinitely old and the ratio of mutation rate to recombination rate is 1). These results indicate that, despite the short ancestral segment, there may be a detectable signal from shared neutral SNPs within it. Moreover, assuming that selection acts on a single site, the targets will be very well delimited, yielding only a handful of possible SNPs to follow up with functional assays.

**LD patterns of shared SNPs in a larger sample**

Given that in genome scans for selection, the targets are not known *a priori*, it is helpful to consider what to expect in a sample of more than four chromosomes, where we do not condition on observing the balanced polymorphism. If $n$ chromosomes are sampled at random from each species, the probability of capturing both alleles at the selected polymorphism in both species is $(1-p^n-q^n)^2$. This probability obviously increases with sample size and is maximized at $p=q=0.5$ for any given sample size.

Assuming that the trans-species polymorphism is observed in the sample, the number of neutral shared SNPs is expected to increase slightly with sample size. For a sample of more than four chromosomes, the ancestral segment can be defined as the distance at which at least one lineage from each allelic class in each species experiences no recombination in stages I and II. This is the segment in which neutral shared SNPs can be found, and its expected length will increase slightly with sample size, because larger samples are less sensitive to chance recombination events very close to the selected site during stage I. However, because stage I is much shorter than stage II (on average less than $4N_e$ generations compared to $\sim T$ generations), this effect will be negligible for the parameters considered (see Table S2).

Increasing the sample size also affects the LD levels between the selected and neutral shared SNPs. In a sample of four chromosomes, if both neutral shared SNP and the selected SNP are present, they will be in perfect LD in both species (*i.e.*, there will be only two haplotypes and $r^2=1$). This is not the case for larger sample sizes, as becomes clear if we consider the ancestral segment as divided into two parts. In the part that is adjacent to the selected site, no recombination occurs on any lineage in both stages I and II, and thus the gene tree topology is exactly the same as for the selected site. Therefore, any shared neutral SNPs (that arise from mutations in stage III) will be in perfect LD with the selected one at present, as is the case for the entire ancestral segment for a sample of four chromosomes. The other part of the ancestral segment is farther from the selected site and defined by there being recombination on some lineages in state I but no recombination in state II. The genealogy of the sample does not cluster by species, nor does it cluster entirely according to the selected allele. Thus, neutral shared SNPs in this part are in imperfect LD with the selected one (*i.e.*, there will be more than two haplotypes). In general, increasing the sample size has a slight influence on the total length of the ancestral segment and moves the boundary between the two parts closer to the selected site, leading to a lower fraction of neutral shared SNPs in perfect LD with the selected one. It follows that increasing the sample size will reduce the expected LD between the selected site and neutral shared SNPs at any given genetic distance.

The level of LD between the selected site and a neutral shared SNP at a distance of *d* Morgans can be thought of as follows. For the neutral site to segregate in both species, it must have arisen in the ancestral population and remained polymorphic in both species in stage I and II. Therefore, when focusing on one species, at the end of stage I (when there are only two lineages, one in each allelic class), one allele (*B1*) at the neutral site must be completely linked to *A1* and another allele (*B2*) completely linked to *A2*. If follows that whether an *A1* lineage at present carries allele *B1* or *B2* depends on whether it remained in the *A1* class or migrated to the *A2* class during stage I (and *vice versa*). Therefore, the LD between the selected and the neutral shared SNPs under consideration reflects the correlation between the lineages' allelic identities at the beginning and the end of stage I.

This correlation can be thought of in terms of the probability that a given lineage does not switch allelic identity during stage I; we denote this probability for a sample of $n$ lineages carrying the same allele by $R_n$. In the Supporting Information, we derive an approximation for $R_n$ and show how it relates to commonly used summaries of pairwise linkage disequilibrium. The approximation under-estimates the extent of LD, but nonetheless helps to understand why the LD levels between the selected and neutral sites are expected to be high even in a large sample, as confirmed by simulation (Figure S2 and Figure S3). In conclusion, if a neutral shared SNP is observed in a sample of more than four chromosomes, it is expected to be in strong but not necessarily in perfect LD with the selected site (Figure 5).

**Simulations of neutral recurrent mutations**

To provide clear-cut evidence for ancient balancing selection, the signals associated with shared SNPs need to be highly unlikely to occur under neutrality. One way that shared SNPs could arise under neutrality is by incomplete lineage sorting (*i.e.*, neutral shared SNPs identical by descent). As noted, when $T$ is much greater than $N_e$, the probability of neutral trans-species polymorphisms is negligible. Moreover, fluctuations in population size will, if anything, tend to *decrease* the probability of neutral trans-species polymorphisms, because bottlenecks would increase the chance of neutral polymorphisms being lost.

For species that are diverged enough, the vast majority of shared SNPs will arise not by identity by descent but by recurrent mutation, *i.e.*, be identical only by state (ASTHANA *et al.* 2005; HODGKINSON *et al.* 2009). To distinguish trans-species SNPs from those shared identical by state, we have to consider their footprints at sites nearby. Notably, while trans-species polymorphisms should be accompanied by neutral shared SNP(s) in LD within the ancestral segment, this combination of signals should be highly unlikely to be generated by neutral recurrent mutations alone.

For illustration, we consider a model with recurrent mutations for humans and Western chimpanzees, incorporating differences in mutation rates between CpG and non-CpG sites (as in LEFFLER *et al.* 2013), because heterogeneity in the mutation rate can greatly increase the rate of recurrent mutations, and, in primates, CpG sites have a much higher mutation rate than non-CpGs (cf. HODGKINSON and EYRE-WALKER 2011) and constitute the majority of recurrent mutations between humans and chimpanzees (LEFFLER *et al.* 2013). We simulate diversity patterns in 10,000 replicates of 100 kb segments (together equivalent to one third of the human genome of $3 \times 10^9$ bps). Within each 100 kb segment, CpGs are randomly placed by sampling the spacing distances from an exponential distribution with a mean of 100bp (~1% dinucleotides in human genome are CpGs). We also consider two demographic scenarios plausible for human populations and Western chimpanzees: one with constant effective population sizes and the other with bottlenecks in both species (see Table S3 for details).

Under the first scenario, only 2% of the segments contain shared SNP pairs within 400 bps and only 7% of these pairs are in significant LD in both species, with the same phase (at the 5% level, using a $\chi^2$-test). Moreover, no replicate contains more than one pair of shared SNPs within 400 bps. As expected, the demographic scenario with bottlenecks (Lı and Durbin 2011; Prado-Martinez *et al.* 2013) leads to even fewer shared SNP pairs: 0.9% replicates contained pairs within 400 bps, of which only 5.6% pairs were in significant LD in both species with the same phase.

In addition, LD patterns among shared SNPs differ for recurrent mutations and under a balanced trans-species polymorphism (Figure 5). In the balancing selection case, $r^2$ between the selected and neutral sites is expected to be near 1 within the expected length of the ancestral segment and to decrease with genetic distance, as expected from our analytic results. In contrast, under recurrent mutations, the $r^2$ between shared SNPs within the same scale of genetic distance are usually quite low (on average around 0.2 for a sample of 50 chromosomes). The reason is that the genealogies underlying two SNPs shared due to recurrent mutations in such close proximity are almost always the same, so the LD between them does not decay with genetic distance, and instead its value mostly depends on the branch on which the mutations arose.

In summary, we expect the combined signatures of trans-species SNPs to be highly specific. Given plausible demographic models, recombination and mutation rates, we estimate that, throughout the genome, neutral recurrent mutations will generate only ~42 shared SNP pairs within 400 bp, in significant LD in both species and with the same alleles associated (given a genome size of $3 \times 10^9$ bp divided into

100 kb windows and given that 2% of windows have SNP pairs, 7% of which are in LD). While this estimate should not be taken too literally, as our simulations ignore some genomic features that may be salient (*e.g.*, variation in recombination rates and heterogeneity in base composition), it is clear from the simulations that considering the number of shared SNPs within a short distance and the LD patterns among them leads to signatures of trans-species polymorphisms that are highly unlikely to be generated by neutral, recurrent mutations alone.

DISCUSSION

Upcoming resequencing data for numerous species provides an unprecedented opportunity to carry out powerful genome-wide scans for trans-species balanced polymorphisms (WIUF et al. 2004; BUBB et al. 2006). In what follows, we consider the implications of our modeling for the design and interpretation of such scans.

**Interpreting existing examples**

A recent study showed that the A/B polymorphism underlying the ABO blood groups shared among many primates is identical by descent in apes and Old World monkeys (SEGUREL et al. 2012). In particular, it is shared between humans and gibbons identically by descent, indicating that the polymorphism originated before the species split of ~19 million years ago (SEGUREL et al. 2012). Strong evidence for identity by descent includes the lack of fixed difference between species and the presence of a second shared (non-synonymous) SNP between humans and gibbons about 100 bps away from the A/B polymorphism, which is in strong LD with the same phase in both species. This pattern of shared SNPs is consistent with our expectations under a trans-species polymorphism: at the genome average recombination rate, an ancestral segment shared between humans and gibbons has an expected length of 35 bps on each side of the selected site, with an 95%-tile of 150 bps. Therefore, this example conforms with the combination of footprints that is expected to be highly specific to a trans-species balanced polymorphism.

A number of additional targets of ancient balancing selection were recently reported in a study of humans and chimpanzees (Leffler *et al.* 2013). In six cases, the diversity levels in humans are comparable with the genome average divergence between humans and chimpanzees; furthermore, the phylogenetic trees around those shared SNPs had the orange topology, providing compelling evidence that the polymorphisms are shared between species identical by descent. Among them, two cases exhibited polymorphism patterns indicative of more complex scenarios of trans-species balanced polymorphisms. A non-coding region (closest to the gene *FREM3*) contains 13 shared SNPs, all at non-CpG sites, which span 6.7 kb and are in almost perfect LD with each other. This region is much longer than expected given the average recombination rate (upper 95%-tile of ~800 bps). One simple explanation is that the recombination rate in this region has been low in both lineages since the split between species, consistent with the low recombination rate estimated from genetic map by International HapMap Project (Frazer *et al.* 2007). An alternative is that recombination between the two haplotypes is selected against, as might be the case if there are two or more sites under selection, with epistasis among them (Kelly and Wade 2000). The large number of shared SNPs in the region nearest *FREM3* (at least 13 SNPs within 6.7 kb) further suggests that the balanced polymorphism is old. Consistent with this, three of these shared SNPs are also segregating in both among 27 gorillas and among ten orangutans, when orangutans split from humans at least 11 Myr ago (Prado-Martinez *et al.* 2013).

A second interesting case is a 1.8 kb region in the first intron of the gene *IGFBP7*, where eight shared SNPs fall into three LD clusters. The first two clusters, which together span ~870 bps, are in almost perfect LD with each other but with the opposite phase in the two species. The switch in phase is also seen at *ABO* (Segurel *et al.* 2012). For *IGFBP7* (but not *ABO*) it is unlikely to be due to recurrent mutations in stage II of the genealogy of a trans-species polymorphism (Figure S4A), because that explanation requires at least two recurrent mutations; more likely are complex recombination events in stage II (Figure S4B). A second pattern of interest is the existence of a third LD cluster (~200 bps in span and ~600 bps from the first two) in high LD with both the other clusters in chimpanzees but not in LD with either in humans. This configuration suggests that distinct balanced polymorphisms led to the first two and the third clusters, with little or no epistasis between them. Moreover, there are two diversity peaks in the region: one underlying cluster one and two and the other underlying cluster three. The high density of shared SNPs in the first two clusters again suggests the balanced polymorphisms might be old, which is supported by the fact that one shared SNP in cluster one is found to be shared by both gorillas and orangutans (Prado-Martinez *et al.* 2013). Overall, among the six strongest candidate regions, the number of shared SNPs is predictive of whether SNPs are shared with other species, consistent with our expectation that older balanced polymorphism should lead to a higher density of shared neutral SNPs. More generally, these examples illustrate how our results can also help to understand more complicated cases of balancing selection than those considered in our simple model.

**Informing the design and interpretation of future studies**

**Choice of species:** Our analyses highlight the split time, $T$, as a key parameter determining the suitability of a pair of species. $T$ needs to be much larger than $N_e$ to avoid being swamped by false positives due to incomplete lineage sorting. For a simple demographic model with random mating and constant population sizes, the probability of neutral trans-species polymorphism is on the order of $e^{-T/N_e}$, which becomes negligible when $T/N_e$ is sufficiently large. At the same time, $T$ needs to be smaller than $1/r$, or the ancestral segment will be too short to contain signatures that distinguish between balancing selection and neutral, recurrent mutations.

It follows that species with large effective population sizes are less appropriate. As we have shown, for *D. simulans* and *D. melanogaster*, and more generally for species with $N_e$ on the order of $10^6$ and recombination rate on the order of $10^{-8}$ per bp per generation (COMERON *et al.* 2012), the ancestral segment around a trans-species polymorphism is expected to be only a few base pairs long. In practice, even if there were additional shared SNPs near the selected one, it would be hard to align sequencing reads with so many polymorphic sites in close proximity, and the trans-species polymorphism might be missed.

In contrast, mammalian species usually have small $N_e$ (LEFFLER *et al.*), so some of mammalian species pairs should be well suited for scans for trans-species polymorphisms. Humans and chimpanzees are a good example: there should be no neutral trans-species polymorphisms in the genome and the ancestral segment is expected to be sufficiently long to contain a combination of signals that is highly specific, including: 1) no fixed differences, 2) additional neutral shared SNP(s) and 3) high LD among these SNPs. Another example of a suitable species pair is rhesus macaques and baboons, as they are thought to have similar demographic parameters to humans and chimpanzees ($T \approx 470,000$ generations and $N_e \approx 36,000$) (Perry *et al.* 2012).

**Specificity:** For species pairs that meet the above criteria, our analyses show that a combination of footprints in the ancestral segment - the lack of fixed differences and the presence of two or more shared SNPs in short distance with high LD among them - should provide clear evidence of a balanced trans-species polymorphism. Indeed, under simple assumptions at least, they are highly unlikely to be generated by neutral, recurrent mutations or by neutral incomplete lineage sorting.

However, the probability of neutral trans-species polymorphisms due to incomplete lineage sorting is affected by demography. While population bottlenecks will only decrease this probability, population structure and admixture increase the mean and variance of the neutral coalescent time, and thus could make neutral trans-species polymorphism more likely. In humans, in particular, the recent structure thought to have existed should have little effect. For example, when the gene flow from Denisova and Neandertal into modern humans is considered, the probability of trans-species neutral polymorphism between humans and chimpanzees for those introgressed regions would increase by only about three-fold (assuming plausible parameter values: $N_e$=10,000, a generation time of 20 years and a split time of 440 Kya between modern humans and these archaic humans (Green *et al.* 2010). Thus, even with such archaic admixture into modern humans, incomplete lineage sorting with chimpanzees is highly unlikely by chance.

In practice, another source of false positives in scans for ancient balancing selection is spurious shared SNPs generated by bioinformatic errors, among which a major concern is the misalignment of sequencing reads from paralogs shared by the two species. This problem can be alleviated by applying stringent bioinformatic filters (criteria on Hardy-Wernberg equilibrium, mapping, coverage, etc.), or by performing experimental validation (LEFFLER *et al.* 2013).

**Sensitivity:** While we assume that the balanced polymorphism arose after balancing selection pressure is established (*i.e.*, selection on *de novo* mutation), it could also arise from standing variation. In the latter case, our results would hold if we substitute the age of the balanced polymorphism $T_S$ by the age of the selection pressure. More generally, our model of balancing selection, albeit simple, highlights the age of the balanced polymorphism (or the age of the selection pressure) as an important parameter that determines the expected number of hitchhiking shared SNPs. Therefore, scans for balancing selection should have high power in detecting older cases.

We also assume no subsequent mutation at the selected site after the first emergence of the balanced polymorphism, thus excluding the possibility of allelic turnover at the selected site (*i.e.*, the replacement of one selected allele by another that arose independently). While allelic turnover might be modest when the mutational target size is small (because the time scale for allelic turnover is very large (TAKAHATA 1990), more generally, cases of long-lived balancing selection could be missed due to allelic turnover. In addition, neutral recurrent mutations on the genealogy of a trans-species polymorphism can also impact power. In stage II, occurrences of the same mutations on the two lineages in the same species lead to fixed differences between species, which are not expected in our model; on the other hand, independent recurrent mutations in the two species give rise to shared SNPs that are identical by state in addition to those identical by descent. These additional shared SNPs can either strengthen or blur the signals of trans-species polymorphism, depending on the LD between them and the selected SNP (WIUF *et al.* 2004).

Another process that can obscure the footprints of trans-species balanced polymorphism is gene conversion. Gene conversion can both decrease the number of shared SNPs and lead to fixed difference between species when it takes place in stage II, and it tends to decrease the LD among shared SNPs when occurring in stage I. Since both recurrent mutations and gene conversion can lead to fixed differences between species, one approach might be to put less weight on the absence of fixed difference in scans of ancient balanced polymorphism and instead to focus on detecting additional shared SNPs in high LD; nonetheless, when observed, the lack of fixed difference is an additional line of evidence in support for a shared polymorphism being trans-species.

Lastly, we made the simplifying assumption of a constant recombination rate per base pair across sites and over time, ignoring, in apes for example, the potential effects of hotspots and their rapid evolution (MYERS *et al.* 2010). Uncertainty about the recombination landscape around a trans-species polymorphism impacts the accuracy of the predicted distribution of ancestral segment length. However, given that the time scale for trans-species polymorphism is so long and that genetic landscape exhibits rapid turnover in the fine scale but appears fairly constant on large scale (of 1Mb), it seems plausible that the average recombination rate at present over a ~1Mb region could reflect the long term average of any small region within it; if so, our approximation for the length of the ancestral segment should still hold.

In summary, for appropriate pairs of species, genome-wide scans for trans-species polymorphism using these footprints should have reasonably high power and extremely low false positive rates, even when considering a wider set of assumptions than explicitly modeled. We emphasize, however, that this does not imply the false discovery rate (FDR) will be low, because the FDR depends on the ratio of the number of regions truly under balancing selection to the number of neutral regions in which these footprints are observed. Thus, even if these footprints are highly specific, if long-lived balancing selection is very rare, true targets may be only a small proportion of candidate regions (*e.g.*, the FDR was estimated to be ~75% in LEFFLER *et al.* 2013).

**Future directions**

We focus on the ancestral segment that has the same genealogical structure as the trans-species polymorphism. However, if the mutation rate is low or the balanced polymorphism does not much predate the species split, there may not be a neutral shared SNP in the ancestral segment. Moreover, sites outside the ancestral segment also carry footprints of balancing selection. In particular, Figure 1 reveals a specific succession of topologies surrounding a trans-species polymorphism: the ancestral segment is always flanked by regions with yellow genealogical topology, an intermediate structure between allele–clustering (orange) and species-clustering patterns (green and blue). By contrast, the flanking sequences of a shared SNP identical by state rarely have the yellow genealogical structure. This observation suggests that methods that explicitly infer underlying genealogical structure from DNA sequence data could leverage additional information about the presence of a trans-species polymorphism. Recently, several modeling frameworks have been developed for statistical inference of local gene genealogies (LI and STEPHENS 2003; MCVEAN and CARDIN 2005; PAUL and SONG 2010; RASMUSSEN and SIEPEL 2013). Such methods could potentially be modified for detection of trans-species polymorphisms, helping to identify additional targets of ancient balancing selection in humans and chimpanzees, as well as in other species. Since there is already high power to detect ancient balanced polymorphisms using the footprints characterized here, we expect these extensions to be most useful for cases where there was no neutral SNP or the ancestral segment happened to be short but the flanking genealogies are highly informative.

## ACKNOWLEDGMENTS

We thank Dick Hudson, Amir Kermany, Laure Segurel and other members of the PPS for helpful discussions.

A

**The selected site under balancing selection**

**Linked neutral sites**

**Unlinked neutral sites**

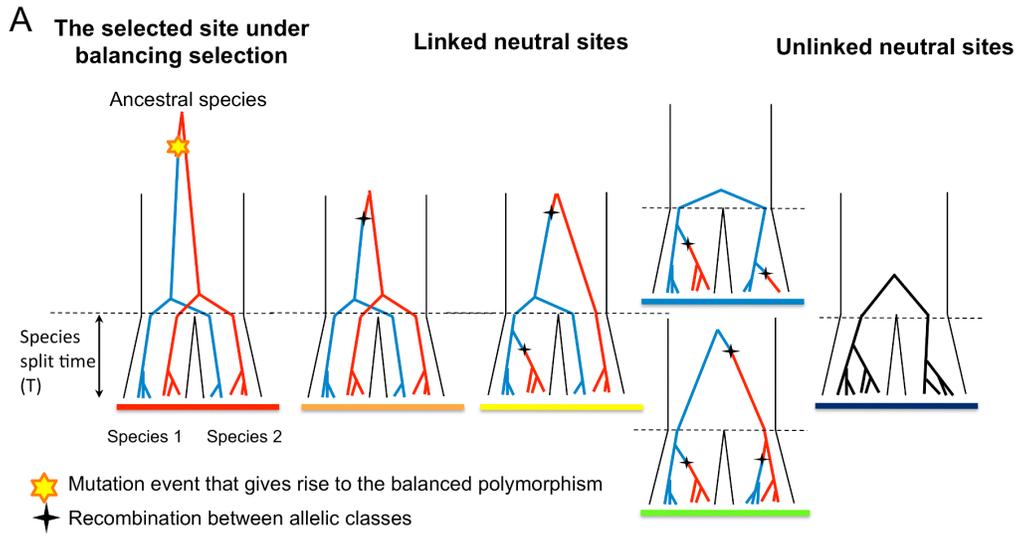

B

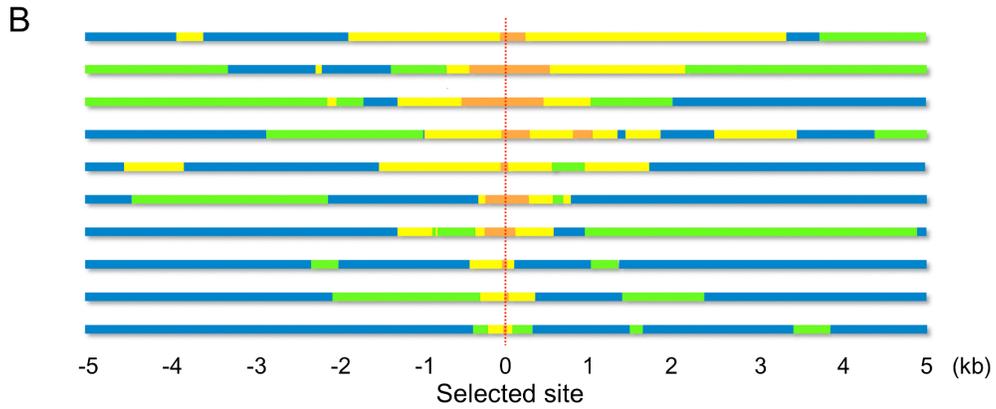

C

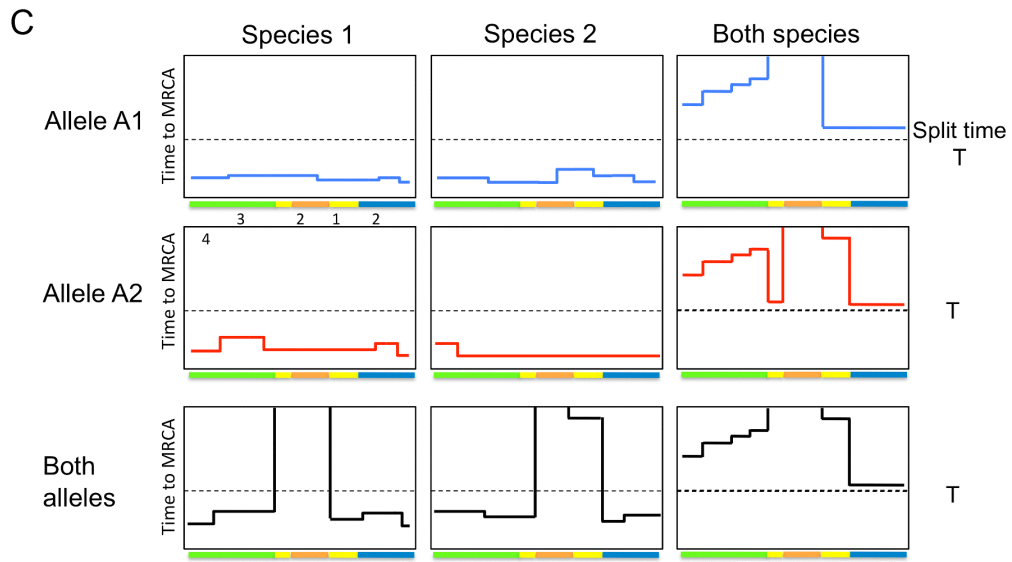

**Figure 1. Sites linked to a balanced trans-species polymorphism have unusual genealogies.** (A) The trees are ordered by the distance from the selected site. Blue and red lines represent lineages from the two allelic classes respectively. (B) The state of each segment in ten simulation replicates. Each bar represents a 10 kb region centered on a trans-species balanced polymorphism (red dotted line). The color of the bar indicates the genealogical state of each segment (same as in A). Parameters were chosen to be plausible for humans and chimpanzees: $N$=10000, $N_a$=50000, $T$=160,000, $p$=0.5 and r=1.25cM/Mb (see Methods). (C) Summary of the coalescent time for a single realization of a segment carrying a balanced trans-species polymorphism. The sample consists of 20 lineages in total, five from each allelic class in each species. Each plot shows the time to the most recent common ancestor (MRCA) for a specific subset of the 20 lineages indicated on the top and left. The selected site is located in the center of the segment.

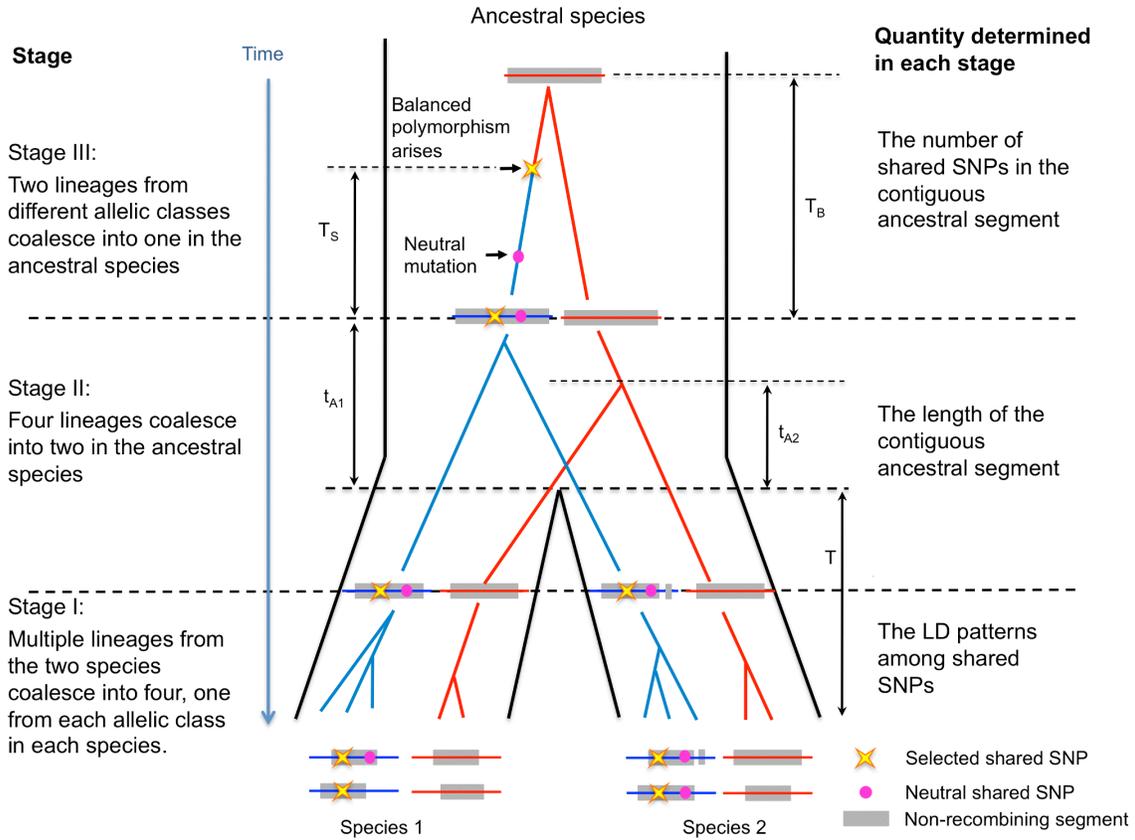

**Figure 2. The genealogy of the contiguous ancestral segment.** The proportion of each stage has been distorted for illustration purpose (for example, stage I should be much shorter compared to stage II). Stage I also has slight influence on the length of the contiguous ancestral segment. See main text for the meaning of symbols.

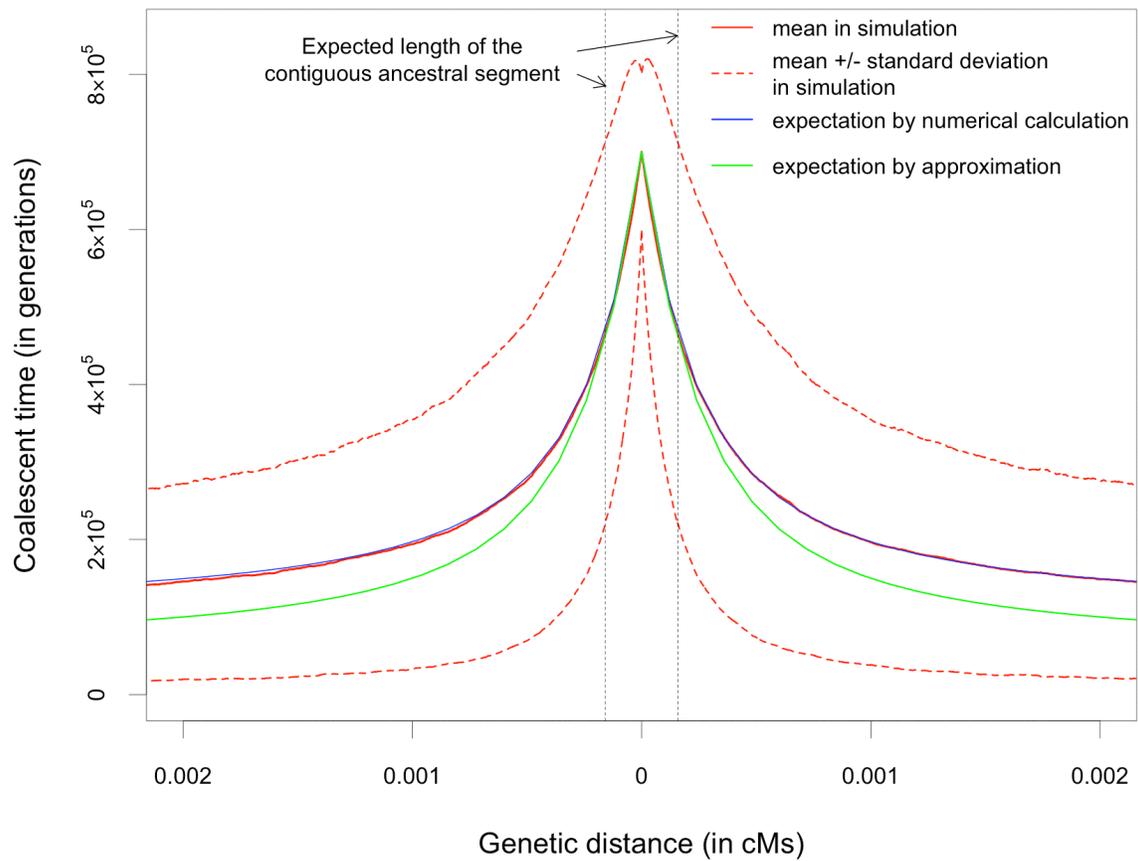

**Figure 3. Expected coalescent time between two lineages carrying different alleles.** Parameters are chosen to be plausible for the ancestral population for humans and chimpanzees: $N_a$=50,000, $p$=0.5 and $T_S$=600,000 generations (see Methods). The mean and standard deviation of the simulation results were obtained from 10,000 replicates.

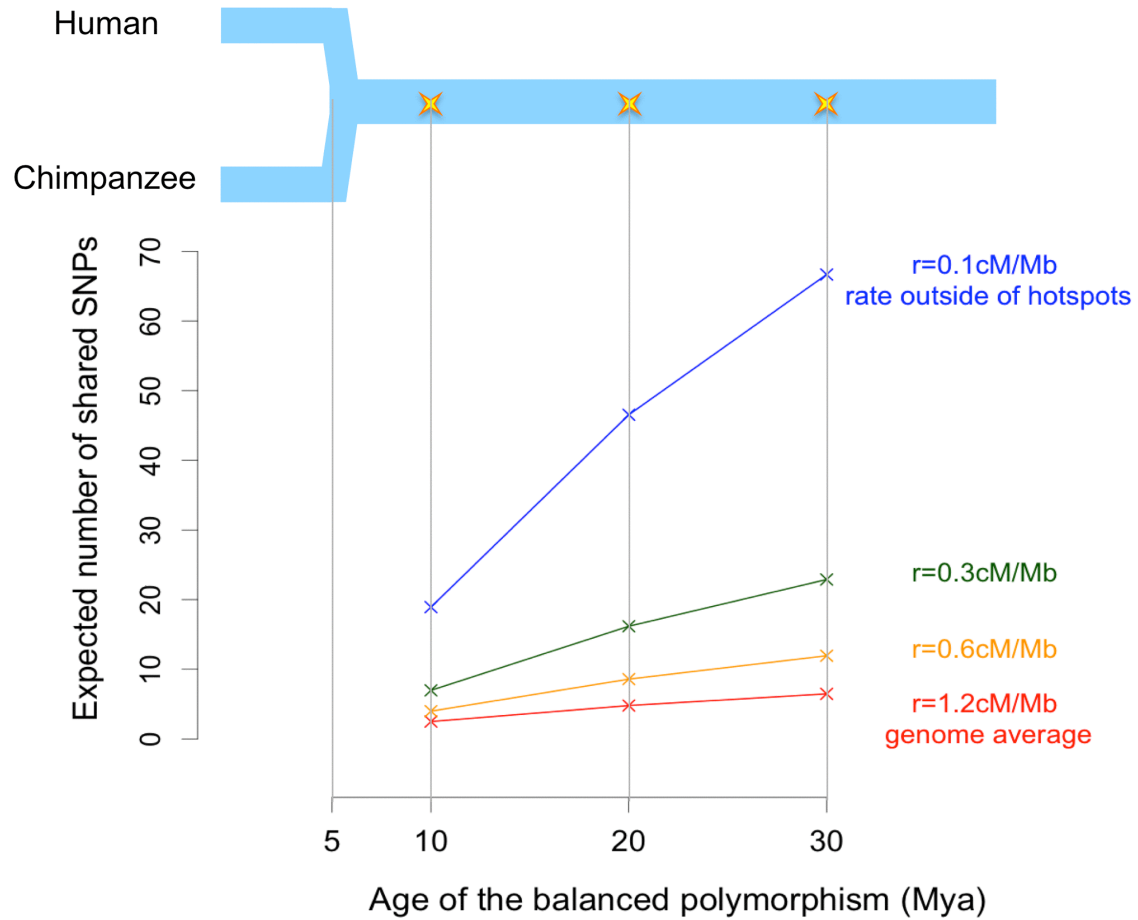

**Figure 4. The expected number of shared SNPs in the contiguous ancestral segment.** The upper panel is a schematic diagram of the demographic history of humans and chimpanzees, which merged into an ancestral species five million years ago (Mya). The star indicates the age of the balanced polymorphism. The lower panel shows the expected number of shared SNPs (including the selected one) in the two-sided contiguous ancestral segment. Note that the age of the balanced polymorphism here is measured from present, which is the sum of $T_S$ and the lengths of stages I and II.

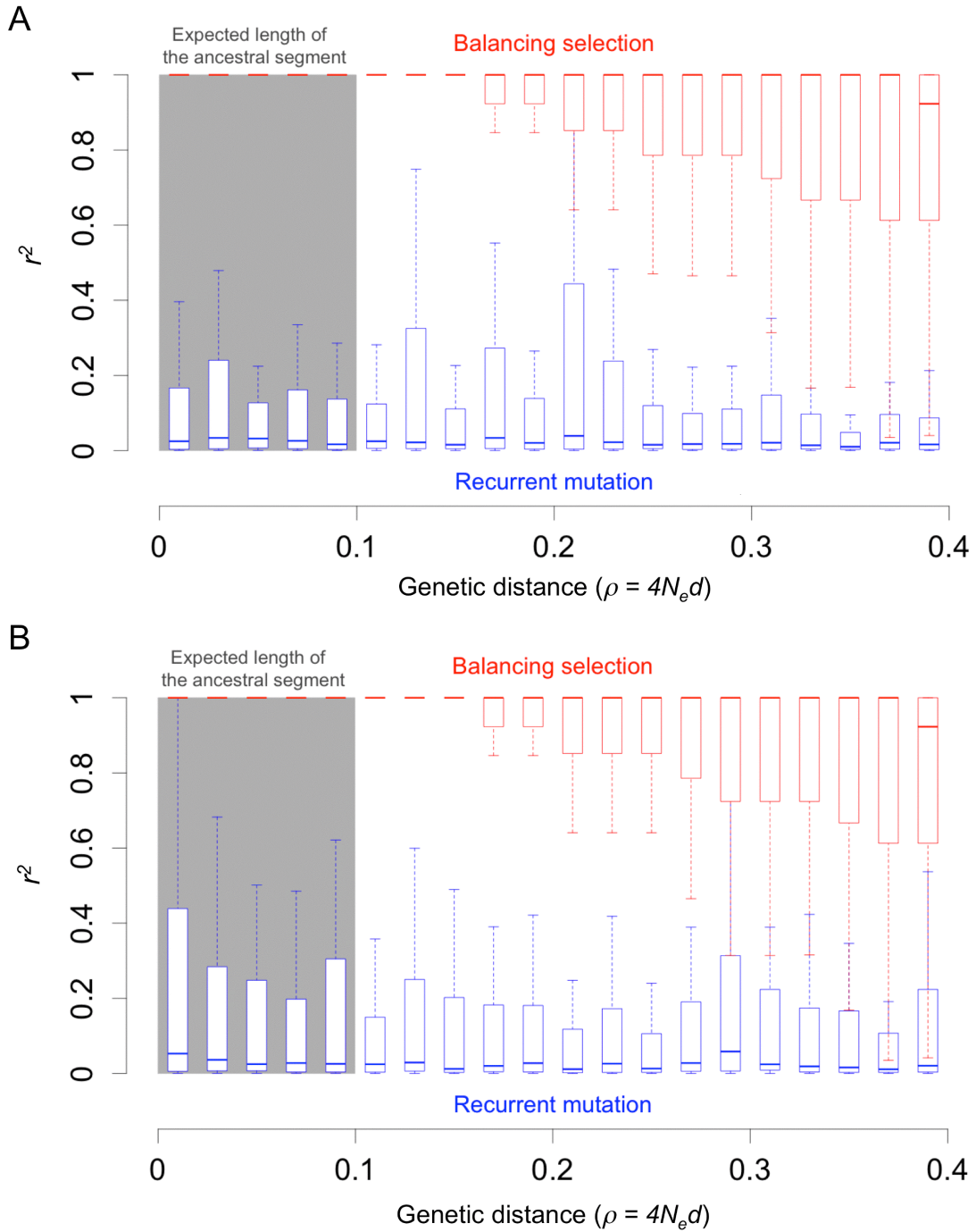

**Figure 5. LD between shared SNPs generated by balancing selection or by recurrent mutations.**
50 chromosomes were sampled from each species in both scenarios. We assumed an allele frequency of 0.5 for the scenario of balancing selection and sampled an equal number of chromosomes from each allelic class (see Methods for details of the simulations). Panels (A) and (B) show the distribution of $r^2$ in the two species respectively.

Supporting Information for

**Footprints of ancient balanced polymorphisms in genetic variation data**

**This file includes:**

**Methods**

I. Why disjoint ancestral segments are unlikely for $T \gg N_e$

II. Derivation of the distribution of the length of the segment with orange or yellow topologies

III. Derivation of $R_n$ and its relation with commonly used summaries of pairwise LD

**Other supporting information for this manuscript includes the following:**

*Mathematica* notebook for the calculation of the expected coalescent time

**I. Why disjoint ancestral segments are unlikely**

Consider four lineages in stage II. Without loss of generality, we assume that the first recombination event in the genealogy occurs on the *A1* lineage in species 1. Now there are two *A2* lineages in species 1 and one *A1* and one *A2* in species 2, so there are three possibilities for the next event: (1) coalescence between the two *A2* lineages in species 1 (2) recombination in species 2 or (3) another recombination in species 1. A disjoint ancestral segment can only be generated under the third scenario.

As discussed in the main text, the genetic distance between the selected site and the location of the first recombination event is on the scale of $1/(2T)$, where $T$ is the split time in generations between the two species, so the recombination rate on any one of the lineages is on the order of $1/(2T)$ per generation, and the coalescent rate between the two *A2* lineages is on the order of $1/(2N_e)$, where $N_e$ is the effective population size for the descendant species. When $T >> N_e$, the rate of coalescence is much higher than that of recombination, so the second event is much more likely to be a coalescent, which leads to a yellow topology; even if the next event is another recombination, there is roughly probability ½ that it takes place in species 2, which will lead to a green or blue topology. Therefore, a disjoint ancestral segment is highly unlikely when $T >> N_e$.

**II. Derivation of the distribution of the length of the segment with orange or yellow topologies**

Without loss of generality, we begin by considering a sample of three chromosomes, two lineages carrying *A1* (one from each species) and one lineage carrying *A2* from species 2. We are interested in the segment where the *A1* lineage from species 2 is more closely related to the *A1* lineage from species 1 than to the *A2* lineage from species 2. The length of this segment is well approximated by the length of the segment in which no recombination event occurs until the two *A1* lineages coalesce. The total coalescent time for the two *A1* lineages is the split time *T* plus the time (*t*) for the two *A1* lineages to coalesce in the ancestral species, where *t* follows an exponential distribution with parameter $2N_a p$, so:

$$f_t(u) = \frac{1}{2N_a p} e^{-\frac{u}{2N_a p}}. \qquad (1)$$

On each side of the selected site, the boundary of the segment of interest is the location of the nearest recombination on any of the three lineages. We denote the length of interest (in genetic distance) by *X* and the genetic distance to the nearest recombination on each lineage by $x_1^{A1}$, $x_2^{A1}$ and $x_2^{A2}$ respectively, so that:

$$X = \min\{x_1^{A1}, x_2^{A1}, x_2^{A2}\}.$$

To calculate the distribution of *x*, we rely on the fact that the distance to the nearest recombination on each lineage is exponentially distributed:

$$x_1^{A1}, x_2^{A1} \sim Exp(q(T+t)),$$

$$x_2^{A2} \sim Exp(p(T+t)).$$

Because $x_1^{A1}, x_2^{A1}$ and $x_2^{A2}$ are independent of each other, the conditional distribution of *X* given *t* is also an exponential distribution:

$$X/t \sim Exp\{(1+q)(T+t)\}. \qquad (2)$$

Combining formula (1) and (2), we obtain the probability density function of $X$ as:

$$f_X(x) = \int_0^\infty f_{X|r}(x \mid u) f(u) du = \int_0^\infty 2(1+q)(T+u) e^{-(1+q)(T+u)x} \frac{1}{2N_a p} e^{-\frac{u}{2N_a p}}) du$$

$$= \frac{(1+q)}{2N_a p} e^{-T(1+q)x} (\frac{T}{A} + \frac{1}{A^2}),$$

where we used $A = x(1+q) + \dfrac{1}{2N_a p}$ to simplify the notation.

### III. Derivation of $R_n$ and its relation with commonly used summaries of pairwise LD

To calculate $R_n$, we make the approximation that each lineage experiences at most one recombination event in stage I. This is a sensible assumption because the length of the ancestral segment (in which shared neutral SNP can be found) is so short that the recombination rate within it is very low, and the probability of two or more recombination events on the same lineage is negligible. This assumption allows us to approximate $R_n$ by the probability that the lineage does not recombine in stage I and to consider the coalescent process in each allelic class separately, because the allelic identity of a lineage switches permanently once it recombines into the other class.

Under this simplifying assumption, $R_n$ can be approximated as follows. In each generation, when $i$ lineages remain in $A1$ class, the number of chromosomes in this allelic class can decrease by one due to a coalescent event with probability $i(i\text{-}1)/(4Np)$ or due to a recombination event with probability $idq$. The probability that a particular lineage recombines and thus leaves the $A1$ class is $dq$ per generation. Therefore, when we focus on one of the $i$ lineages, the probability that it remains in the $A1$ class after the next event happens (i.e., when i-1 lineages remain) is: $1 - \dfrac{dq}{idq + \dfrac{i(i-1)}{4N_e p}} = \dfrac{i-1}{i} \cdot \dfrac{\rho pq + i}{\rho pq + i - 1}$, where $\rho = 4N_e d$ (the population genetic distance). It follows that the probability that the lineage does not switch allele identity during stage I is:

$$R_n = \prod_{i=2}^{n} \frac{i-1}{i} \cdot \frac{\rho pq + i}{\rho pq + i - 1} = \frac{1}{n} \cdot \frac{\rho pq + n}{\rho pq + 1} = \frac{1}{n}(1 + \frac{n-1}{\rho pq + 1}) = \frac{1}{\rho pq + 1}(1 + \frac{\rho pq}{n}). \quad (3)$$

From the simple expression for $R_n$ in equation (3), it can be seen that, as expected, $R_n$ decreases with $n$. The decrease is due to the increase in the expected time to the most recent common ancestor ($T_{\text{MRCA}}$) of the sample of $A1$ lineages, which leads to a higher probability for each lineage to recombine during that time; since the expected $T_{\text{MRCA}}$ approaches a limit ($4N_e p$) when n is large, $R_n$ has a limit when $n$ approaches infinity. Also as expected, $R_n$ decreases with $\rho$. Interestingly, $R_n$ is symmetric for $p$ and $q$ and decreases with their product, so it decays fastest with distance when the allele frequency is $p=q=0.5$.

The approximation in equation (3) is expected to underestimate the probability that a lineage retains its allelic identity in stage I, because $R_n$ does not include the probability that a lineage recombines two (or more) times and thus re-enter the $A1$ class. However, it still performs reasonably well, especially on the length scale of the ancestral segment (see Figure S2 for a comparison to simulation results).

$R_n$ can be used to approximate the expectation of widely used summaries of pairwise LD, such as D, D' and r2 It can be easily shown that these properties of $R_n$ carry over to the widely used summaries of LD. In a sample of n $A1$–carrying lineages at present, each lineage has probability $R_n$ to carry $B1$ allele at the neutral polymorphic site and probability 1- $R_n$ to carry $B2$; therefore, making the (incorrect) simplifying assumption of independence between lineages, the total number of $A1B1$ haplotypes, $n11$, can be approximated by the binomial distribution $Bin(n, R_n)$ and the other ($n - n11$) lineages would be $A1B2$ haplotypes. Similarly, with a sample of $n1$ $A1$ lineages and $n2$ $A2$ lineages, $n1$, $n2$, $R_{n1}$ and $R_{n2}$ can be used to approximate the distribution of the numbers (or frequencies) of the four possible haplotypes, with which $D$, $D'$ and $r^2$ can be calculated. Simulations suggest that this approximation underestimates the extent of LD but confirm that the LD between the selected and neutral sites are expected to be very high even in a large sample (Figure S3).

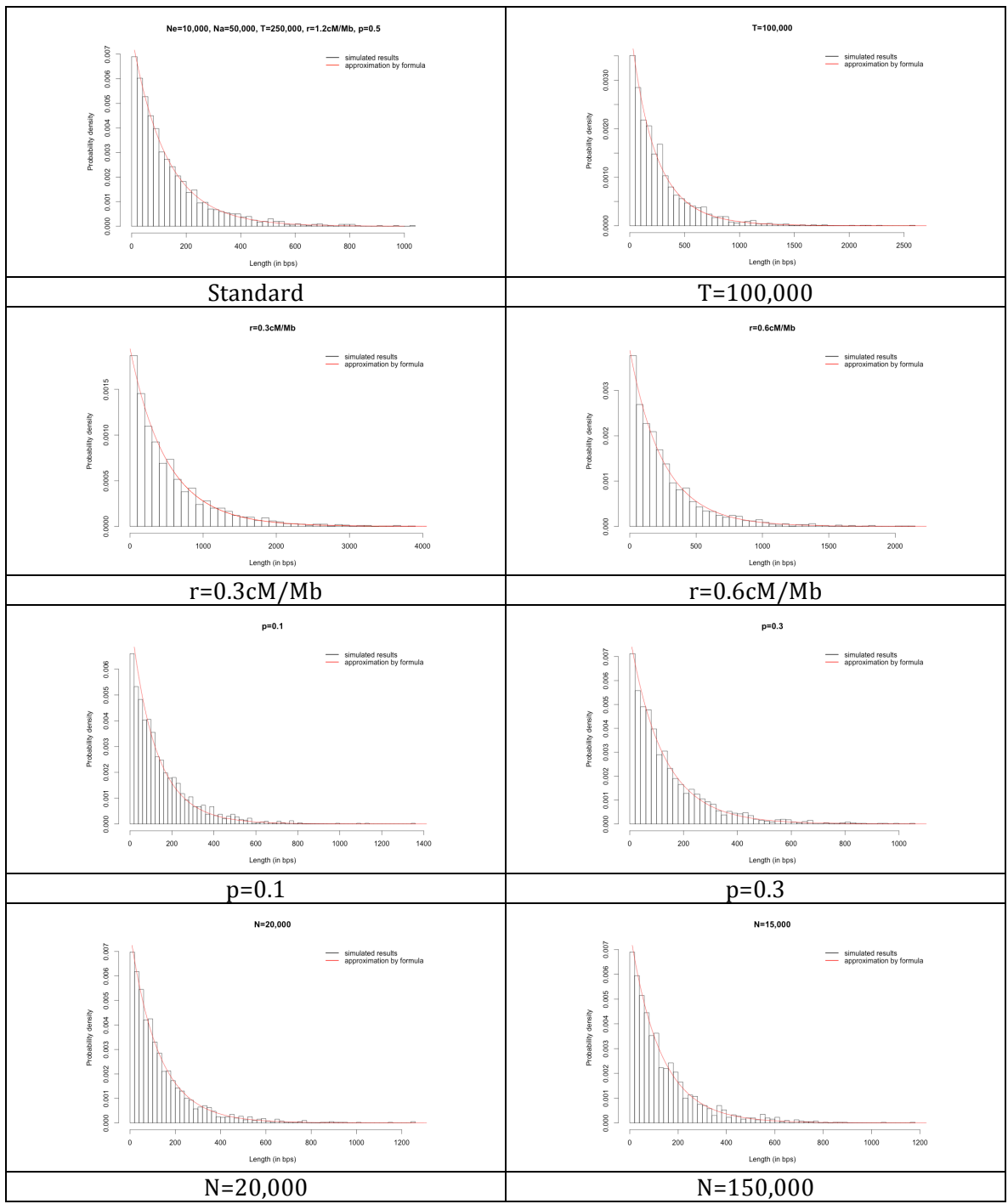

**Figure S1**: Distribution of the length of ancestral segment (simulation vs. approximation). If not otherwise specified, parameters used were: Ne=10,000, Na=50,000, T=250,000, r=1.2cM/Mb and p=0.5, which were chosen to be plausible for humans and chimpanzees. For all combinations of parameters tested, the distribution predicted by the approximation fits the simulation results very well.

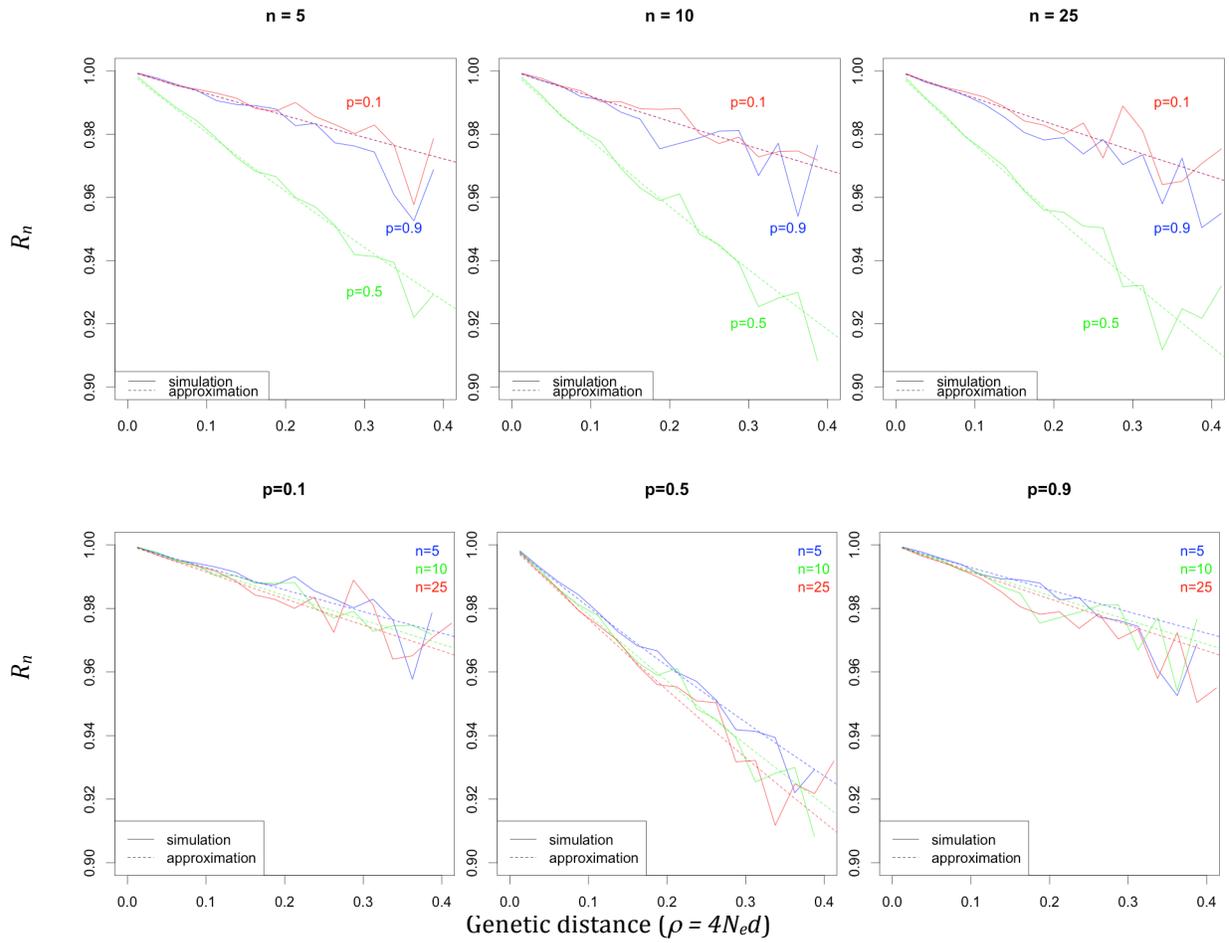

**Figure S2:** $R_n$ calculated from approximation and from simulation. The simulated segment is divided into bins of $6.25\times10^{-5}$ cM, and the mean $R_n$ is calculated from all SNPs in each bin. Only bins with more than 50 data points are shown. See methods for details of the simulations.

**A**

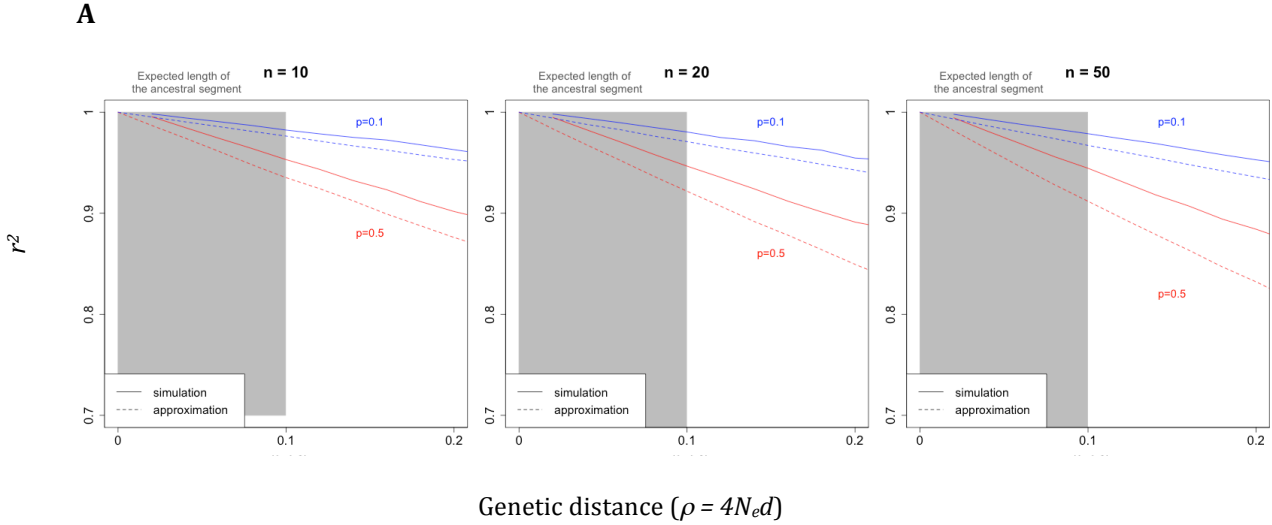

**B**

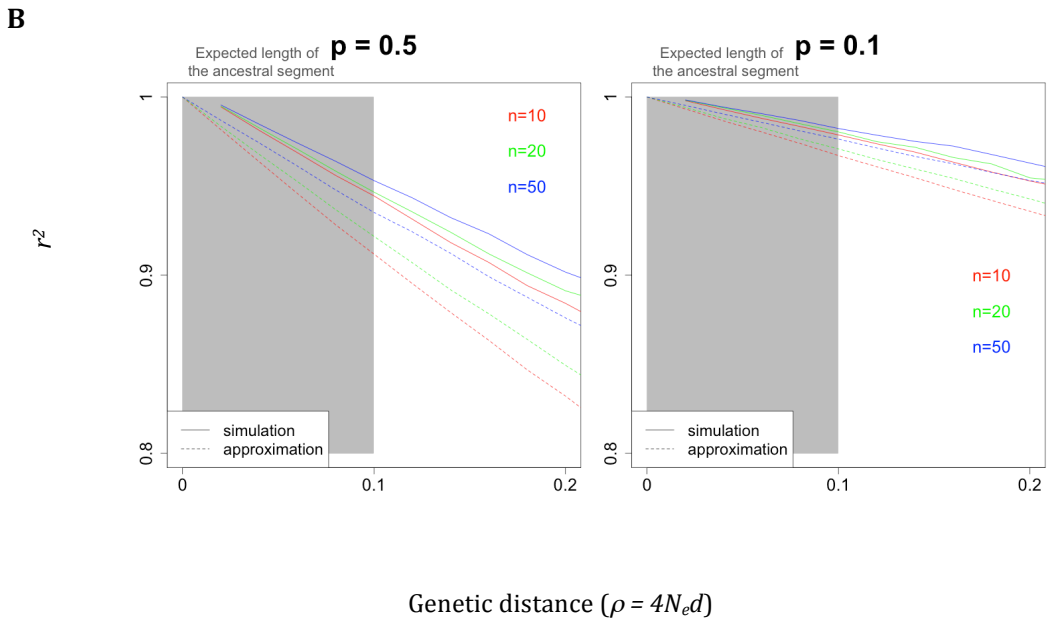

**Figure S3:** Expected *r²* calculated from approximation and from simulation. The simulated segment is divided into bins of 6.25×10⁻⁵ cM, and the mean $R_n$ is calculated from all SNPs in each bin. The expected *r²* from $R_n$, was calculated from 10,000 "virtue simulations": for each simulation, we first sampled $n_1$ *A1* chromosomes from $Bin(50, p)$ and then sampled $n_{11}$ *A1B1* chromosomes from $Bin(n_1, R_{n_1})$ and $n_{22}$ *A2B2* chromosomes from $Bin(n_2, R_{n_2})$, based on which we calculated *r²* and took the average across simulations as the expectation. (A) The mean (or expected) *r²* decreases with sample size. (B) The mean (or expected) *r²* decreases with the product of the allele frequencies.

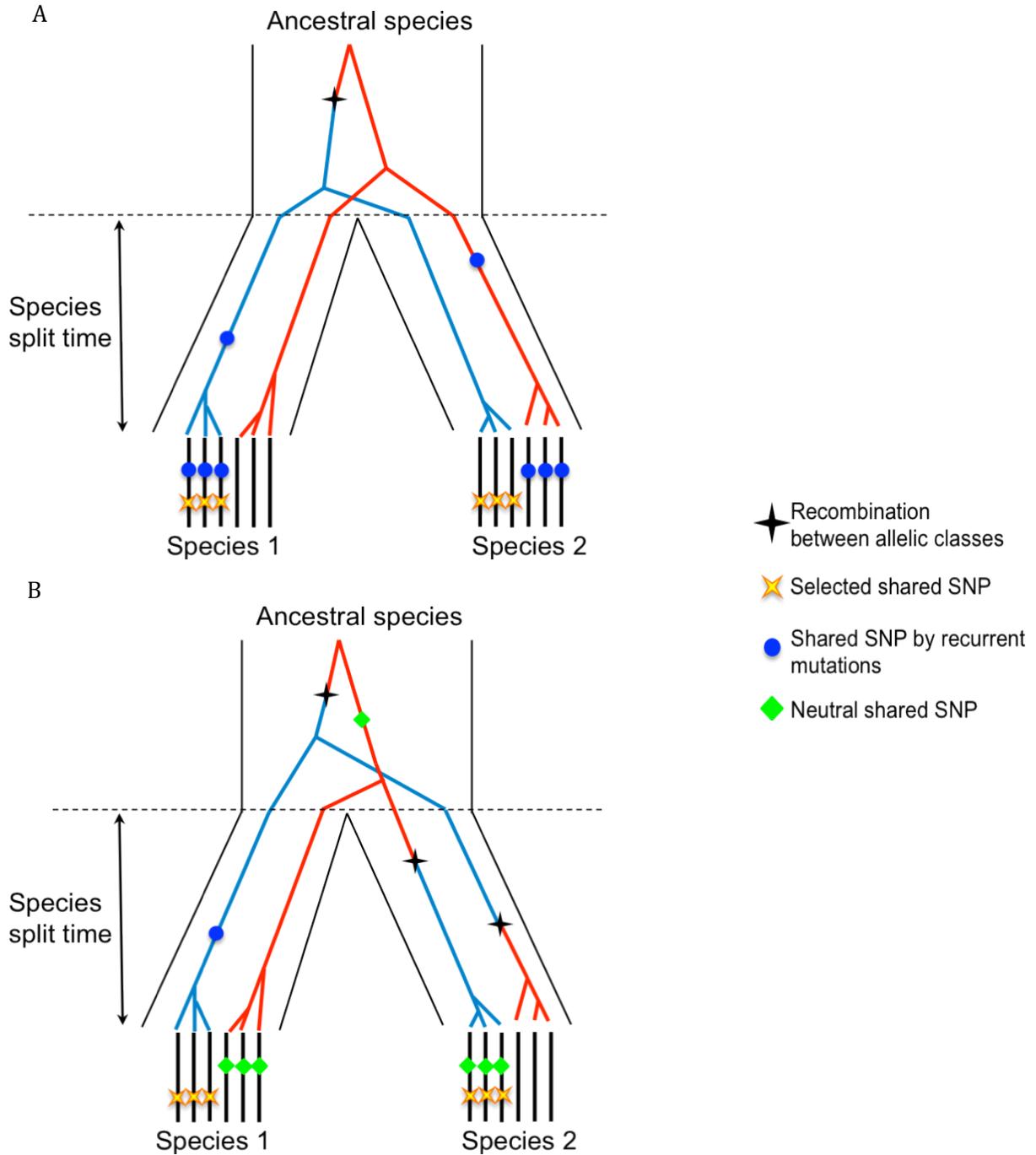

**Figure S4**: Two scenarios that can generate shared polymorphisms in LD but with the opposite phases in the two species. (A) Recurrent mutations on the genealogy of a trans-species polymorphism: independent occurrences of same mutation occur in both species but arise on lineages carrying different selected alleles in the two species. (B) Complex recombination events: two recombination events occur in species 2 before the split, which switch the allele identities of the two lineages in that species.

**Table S1: Summaries of the length of the contiguous ancestral segment (simulation vs. approximation).**

| Split time (*T*) | | Length (in bps) | | | |
|---|---|---|---|---|---|
| | | 1st Qu. | Median | 3rd Qu. | Mean |
| 100,000 | Approximation | 70 | 171 | 352 | 258 |
| | Simulation | 76 | 183 | 361 | 274 |
| 250,000 | Approximation | 38 | 90 | 182 | 131 |
| | Simulation | 39 | 93 | 188 | 137 |
| 500,000 | Approximation | 21 | 51 | 102 | 73.1 |
| | Simulation | 23 | 55 | 107 | 76.9 |

| Recombination rate (*r*) | | Length (in bps) | | | |
|---|---|---|---|---|---|
| | | 1st Qu. | Median | 3rd Qu. | Mean |
| 1.2cM/Mb | Approximation | 38 | 90 | 182 | 131 |
| | Simulation | 39 | 93 | 188 | 137 |
| 0. 6cM/Mb | Approximation | 75 | 180 | 363 | 262 |
| | Simulation | 72 | 179 | 359 | 266 |
| 0. 3cM/Mb | Approximation | 149 | 359 | 725 | 525 |
| | Simulation | 143 | 361 | 763 | 545 |

| Population size ($N_e$) | | Length (in bps) | | | |
|---|---|---|---|---|---|
| | | 1st Qu. | Median | 3rd Qu. | Mean |
| Not involved | Approximation | 38 | 90 | 182 | 131 |
| 10000 | Simulation | 39 | 93 | 188 | 137 |
| 15000 | Simulation | 39 | 94 | 190 | 140 |
| 20000 | Simulation | 38 | 90 | 181 | 136 |

| Allele frequency (*p*) | | Length (in bps) | | | |
|---|---|---|---|---|---|
| | | 1st Qu. | Median | 3rd Qu. | Mean |
| 0.1 | Approximation | 43 | 101 | 204 | 146 |
| | Simulation | 36 | 87 | 177 | 129 |
| 0.2 | Approximation | 42 | 97 | 197 | 142 |
| | Simulation | 37 | 88 | 179 | 130 |
| 0.3 | Approximation | 40 | 94 | 189 | 140 |
| | Simulation | 37 | 89 | 181 | 131 |
| 0.4 | Approximation | 37 | 96 | 194 | 140 |
| | Simulation | 38 | 90 | 181 | 131 |
| 0.5 | Approximation | 39 | 93 | 188 | 137 |
| | Simulation | 38 | 90 | 182 | 131 |

Parameters used were: $N_e$ =10,000, $N_a$=50,000, $T$=250,000, $r$=1.2cM/Mb and $p$=0.5.

**Table S2: Influence of sample size on the number and LD of the neutral shared SNPs.**

| Total sample size (number of chromosomes sampled from each class in each species) | Average number of shared neutral SNPs per replicate (100,000 replicates) | | | |
|---|---|---|---|---|
| | All | SNPs in perfect LD with the selected one in species 1 | SNPs in perfect LD with the selected one in species 2 | SNPs in perfect LD with the selected one in both species |
| 4 (1) | 19.022 | 19.022 (100%) | 19.022 (100%) | 19.022 (100%) |
| 20 (5) | 19.690 | 18.884 (95.9%) | 18.865 (95.8%) | 18.154 (92.2%) |
| 40 (10) | 19.767 | 18.657 (94.4%) | 18.666 (94.4%) | 17.726 (89.7%) |
| 100 (25) | 19.798 | 18.333 (92.6%) | 18.353 (92.6%) | 17.162 (86.7%) |
| 200 (50) | 19.811 | 18.140 (91.6%) | 18.128 (91.5%) | 16.811 (84.9%) |

Same number of chromosomes were sampled from each allelic class in each species. Parameters were chosen to be plausible for humans and chimpanzees: $p$=0.5, $T$=20$N_e$, $N_a$= $N_e$, and $T_{BS}$=400 $N_e$. We assumed an old age of the balanced polymorphism to increase the probability of observing at least one neutral shared SNP in a replicate but this will not affect the LD between neutral and selected shared SNPs (results not shown). When calculating the proportion of neutral SNPs in perfect LD, we picked one neutral shared SNP at random from each replicate to ensure independence.

**Table S3: Parameter values used in simulations of neutral recurrent mutations.**

**Parameters for both demographic models**

| Parameter | Value |
|---|---|
| Mutation rate | $1.8\times10^{-8}$ per generation per base pair |
| Recombination rate | 1.2 cM/Mb per generation (sex-averaged mean recombination rate in human genome) |
| Sample size | 50 chromosomes from each species |
| Segment length | 100 kb |
| Proportion of CpG sites | 2% |

**Demographic model with constant population sizes**

| Parameter | Value |
|---|---|
| Effective population size for humans($N_h$) | 15,800[a] |
| Effective population size for Western chimpanzees ($N_c$) | 11,100[a] |
| Split time ($T$) | 240,000 generations[b] |
| Effective population size for ancestral species ($N_a$) | 50,000 |

[a] Derived from a mutation rate of $1.8\times10^{-8}$ and the observed heterozygosity
[b] Derived from a split time of 6 Myr, assuming a generation time of 25 years

**Demographic model with bottlenecks**

| Parameter | Value |
|---|---|
| Effective population size for humans ($N_h$) | 13,900[c] |
| Effective population size for Western chimpanzees ($N_c$) | 12,500[d] |
| Period of the bottleneck for humans | 28-56 kya[c] |
| Period of the bottleneck for Western chimpanzees | 15-35 kya[d] |
| Effective population size for humans during the bottleneck ($N_h'$) | 2,200 |
| Effective population size for Western chimpanzees during the bottleneck ($N_c'$) | 3,500 |
| Split time ($T$) | 240,000 generations |
| Effective population size for ancestral species ($N_a$) | 50,000 |

[c] Li, H., and R. Durbin, 2011 Inference of human population history from individual whole-genome sequences. Nature 475**:** 493-496.
[d] Prado-Martinez, J., P. H. Sudmant, J. M. Kidd, H. Li, J. L. Kelley *et al.*, 2013 Great ape genetic diversity and population history. Nature 499**:** 471-475.